\documentclass[conference]{IEEEtran}   % TWO COLUMN IEEE

\usepackage{cite}
\usepackage{amsmath,amssymb,amsfonts}
\usepackage{algorithmic}
\usepackage{graphicx}
\usepackage{textcomp}
\usepackage{xcolor}
\usepackage{setspace}
\usepackage{float}
\usepackage{url}

\newcommand{\figwidth}{\columnwidth}   % TWO COLUMN IEEE

\def\BibTeX{{\rm B\kern-.05em{\sc i\kern-.025em b}\kern-.08em
    T\kern-.1667em\lower.7ex\hbox{E}\kern-.125emX}}

\IEEEoverridecommandlockouts
\begin{document}

% -------------------------------------------------------
%  TITLE
% -------------------------------------------------------
\title{VaultxGPU: GPU-Accelerated Blockchain Consensus}

% -------------------------------------------------------
%  AUTHORS
%  [TOGGLE 4/4] -- comment out the block you are NOT using
% -------------------------------------------------------

% TWO COLUMN IEEE -- comment out if using single column
\author{
  \IEEEauthorblockN{Samuel Taiwo Fatunmbi}
  \IEEEauthorblockA{
    \textit{College of Computing} \\
    \textit{Illinois Institute of Technology}\\
    Chicago, USA \\
    sfatumnbi@hawk.illinoistech.edu
  }
  \and
  \IEEEauthorblockN{Om Amit Gandhi}
  \IEEEauthorblockA{
    \textit{College of Computing} \\
    \textit{Illinois Institute of Technology}\\
    Chicago, USA \\
    ogandhi1@hawk.illinoistech.edu
  }
  \and
  \IEEEauthorblockN{Luke Logan}
  \IEEEauthorblockA{
    \textit{College of Computing} \\
    \textit{Illinois Institute of Technology}\\
    Chicago, USA \\
    llogan@illinoistech.edu
  }
}

% SINGLE COLUMN -- comment out if using two column IEEE
% \author{
%   Samuel Taiwo Fatunmbi\\
%   \textit{College of Computing}\\
%   \textit{Illinois Institute of Technology}\\
%   Chicago, USA\\
%   sfatumnbi@hawk.illinoistech.edu
%   \and
%   Om Amit Gandhi\\
%   \textit{College of Computing}\\
%   \textit{Illinois Institute of Technology}\\
%   Chicago, USA\\
%   ogandhi1@hawk.illinoistech.edu
%   \and
%   Luke Logan\\
%   \textit{College of Computing}\\
%   \textit{Illinois Institute of Technology}\\
%   Chicago, USA\\
%   llogan@illinoistech.edu
% }

\maketitle

% -------------------------------------------------------
%  ABSTRACT
% -------------------------------------------------------
\begin{doublespace}
\begin{abstract}
Blockchain consensus mechanisms based on Proof-of-Work consume significant
energy, with Bitcoin alone estimated at approximately 150 TWh per year.
Proof-of-Space reduces this cost by replacing repeated computation with
storage, but plot generation remains bottlenecked by CPU hashing throughput.
Prior work on VaultX demonstrated a high-performance CPU-based Proof-of-Space
plotter using multi-threaded Blake3 hashing, achieving plotting speeds 4 to
50x faster than Chia depending on hardware configuration. In this paper, we
present VaultxGPU, a GPU-accelerated extension of the VaultX plotter that
offloads the Blake3 hashing pipeline to the GPU using custom kernels. We
implement the plotter in both CUDA for NVIDIA hardware and SYCL for AMD and
Intel GPUs, keeping Table 1 entirely in GPU VRAM and fusing the sort and
match stages into a single kernel to minimize data movement. We evaluate
VaultxGPU across K-values 27 through 31 against CPU baselines. Our SYCL GPU
implementation achieves a 59.2x speedup over a single-threaded CPU baseline,
completing a K=31 plot in 45.4 seconds compared to 2688 seconds, and
outperforms even the best 384-thread CPU configuration. These results confirm
that GPU acceleration is the correct direction for scaling Proof-of-Space
plotting beyond what CPU parallelism can achieve.
\end{abstract}

% [TOGGLE] Keywords format
% TWO COLUMN IEEE:
% SINGLE COLUMN:
% \noindent\textbf{Keywords:} Blockchain, Proof-of-Space, GPU computing,
% CUDA, SYCL, Blake3, parallel hashing

% ===============================================================
%  PART 1 — PROBLEM DESCRIPTION
% ===============================================================
\section{Introduction}

Blockchain networks rely on consensus mechanisms to ensure all participants
agree on the state of a distributed ledger. The dominant mechanism,
Proof-of-Work (PoW), requires miners to repeatedly compute cryptographic
hashes in a race to solve an arbitrary puzzle, consuming approximately 150
TWh of electricity per year in the case of Bitcoin alone \cite{b_ccaf},
which is comparable to the annual energy consumption of many mid sized
countries. Proof-of-Space (PoSp) offers a more energy-efficient alternative
by replacing repeated computation with storage. Miners pre-compute and store
large plot files on disk, then respond to challenges by looking up matching
entries rather than hashing on demand, reducing energy consumption by
approximately 1000x \cite{b_chia}. However, the one-time cost of generating
these plot files is itself bottlenecked by CPU hashing throughput, limiting
how quickly a miner can begin participating in the network. Prior work on
VaultX introduced a CPU-based PoSp plotter using multi-threaded Blake3
hashing that significantly outperforms existing implementations
\cite{b_vaultx}, yet even a high-end 192-core CPU costing upwards of
\$10,000 is outpaced by a \$5,000 machine equipped with a single Tesla V100
GPU, a gap that motivates the GPU-accelerated approach presented in this
paper.

\subsection{Challenges}

Accelerating Proof-of-Space plot generation on the GPU introduces several
non-trivial engineering challenges. The first is parallelizing the Blake3
keyed hashing algorithm \cite{b_blake3}, which was designed for sequential
CPU execution. In VaultxGPU, each of the $2^k$ nonces must be processed by
an independent GPU thread, requiring the hash to be computed entirely within
a single thread's register file with no shared state, a fundamental redesign
of the data flow rather than a simple port. The second challenge is
cross-vendor support: NVIDIA and AMD require entirely separate programming
models. In the CUDA implementation \cite{b_cuda}, the 256-bit plot key is
stored in \texttt{\_\_constant\_\_} memory and broadcast to all threads in a
warp at zero memory bandwidth cost, while atomic bucket insertions are handled
implicitly by the runtime. In the SYCL implementation \cite{b_sycl}, constant
memory does not exist as a concept; the key must instead be allocated with
\texttt{malloc\_device} and passed as a pointer argument, introducing
additional memory fetch overhead, and all atomic operations must be declared
explicitly using fully-typed \texttt{sycl::atomic\_ref} objects with defined
memory ordering, scope, and address space. The third challenge is that the
entire Table-1 data structure. All $2^k$ nonce-hash records organized into
$2^{24}$ buckets which must reside in GPU VRAM throughout hashing and
matching, as any intermediate transfer over PCIe would negate the GPU's
throughput advantage. At $K=31$ this alone requires 24.7-GB of VRAM,
pushing against the physical limits of even high-end hardware. Finally, the
sort and match stage runs insertion sort entirely within a single thread per
bucket, a deliberate simplicity tradeoff that becomes an $O(n^2)$ bottleneck
as bucket sizes grow with $K$, contributing to the sharp rise in Sort/Table-2
time from 9--11\% of total pipeline time at $K=27$--$29$ to 34\% at $K=31$.

\subsection{Contributions}

VaultxGPU makes the following contributions to GPU-accelerated Proof-of-Space
plotting. First, we design and implement a custom GPU kernel for Blake3 keyed
hashing \cite{b_blake3} that assigns one thread per nonce across the full
$2^k$ nonce space, replacing the sequential CPU hashing pipeline entirely and
eliminating all inter-thread dependencies by computing each hash entirely
within a single thread's register file. Second, we provide dual backend
implementation. CUDA \cite{b_cuda} targeting NVIDIA hardware via \texttt{nvcc}
and SYCL \cite{b_sycl} targeting AMD and Intel GPUs via Intel oneAPI DPC++,
sharing a common Blake3 kernel in \texttt{blake3\_common.h} while adapting
memory management, atomic semantics, and key storage to each programming
model. Third, we fuse the sort and Table 2 generation stages into a single
GPU kernel that operates entirely within shared memory per bucket, avoiding
any intermediate global memory round-trip between the two stages and launching
one block per bucket across all $2^{24}$ buckets simultaneously. Fourth, we
integrate VaultxGPU into the VaultX Proof-of-Space consensus mechanism
\cite{b_vaultx}, producing plot files that are byte-compatible with the
existing CPU VaultX prover, enabling GPU-generated plots to be searched
directly without any format conversion. Finally, we present a detailed
performance evaluation across K-values 27 through 31, comparing CUDA and
SYCL GPU backends against both a naive single-threaded CPU baseline and the
best 384-thread CPU configuration, and characterize per-stage bottlenecks in
the pipeline across all tested K-values.

% ===============================================================
%  PART 2 — IMPORTANT EXISTING SOLUTIONS
% ===============================================================
\section{Related Work}

The work presented in this paper sits at the intersection of three active
research areas which are energy-efficient blockchain consensus, Proof-of-Space
protocol design, and GPU-accelerated cryptographic hashing. This section
surveys the most relevant prior work in each area and positions VaultxGPU
with respect to existing approaches, highlighting the gap that
GPU-accelerated Proof-of-Space plotting fills.

\subsection{Proof-of-Work and Its Energy Cost}

Proof-of-Work (PoW) consensus, first introduced by Bitcoin \cite{b_bitcoin},
requires network participants to repeatedly compute SHA-256 cryptographic
hashes in search of a hash value below a dynamically adjusted target
threshold. The computational difficulty of this puzzle is calibrated so that
the network collectively finds a valid solution approximately once every ten
minutes, ensuring a predictable block production rate regardless of total
network hash power. While this mechanism provides strong security guarantees
through its economic cost of attack, it achieves security at the direct
expense of energy. As of 2021, the Bitcoin network alone consumed an estimated
143 TWh of electricity annually \cite{b_ccaf}, a figure that places it on par
with the total electricity consumption of countries such as Norway and
Bangladesh. This energy expenditure is structurally wasteful, every hash
computed during a mining round that does not meet the target is discarded
immediately, contributing nothing to the network beyond demonstrating that
work was performed. The introduction of application-specific integrated
circuits (ASICs) optimized for SHA-256 hashing has further concentrated mining
power and driven energy consumption upward, as miners are incentivized to
deploy ever-increasing amounts of specialized hardware to remain competitive.
Numerous alternative consensus mechanisms have been proposed to address this
inefficiency, including Proof-of-Stake \cite{b_pos}, which replaces
computational work with economic stake, and Proof-of-Space \cite{b_chia},
which replaces computation with storage. VaultxGPU builds on the
Proof-of-Space direction, accepting its one-time plot generation cost as a
worthwhile tradeoff against the ongoing energy drain of Proof-of-Work.

\subsection{VaultX CPU Baseline}

\begin{figure}[htbp]
    \centerline{\includegraphics[width=\figwidth]{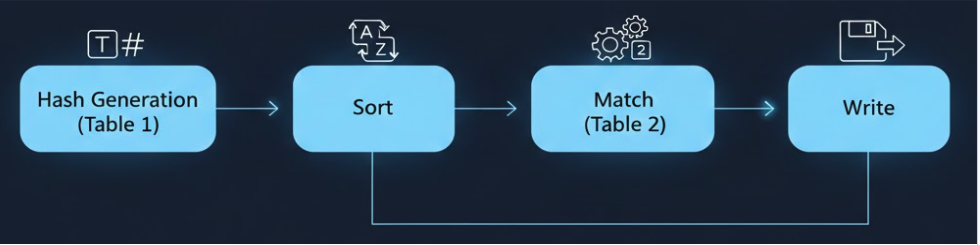}}
    \caption{VaultX CPU plotting pipeline: Hash Generation (Table 1)
    $\rightarrow$ Sort $\rightarrow$ Match (Table 2) $\rightarrow$ Write.}
    \label{fig:cpu_pipeline}
\end{figure}

\begin{figure}[htbp]
    \centerline{\includegraphics[width=\figwidth]{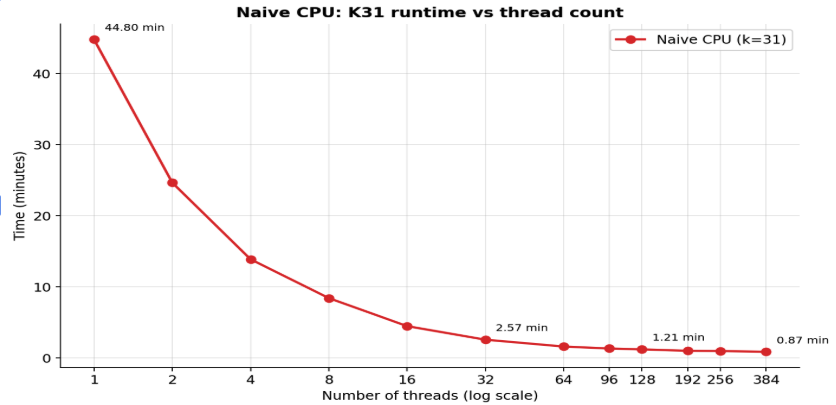}}
    \caption{VaultX CPU K=31 runtime vs.\ thread count across server-grade
    hardware, showing diminishing returns beyond 64 threads.}
    \label{fig:cpu_results}
\end{figure}

The VaultX CPU plotter \cite{b_vaultx}, the direct predecessor to VaultxGPU,
implements a four-stage pipeline as shown in Fig.~\ref{fig:cpu_pipeline}:
hash generation produces Table-1 by hashing all $2^k$ nonces using the Blake3
keyed hash algorithm \cite{b_blake3} and storing nonce-hash pairs into
contiguous memory buckets, followed by an in-place sort within each bucket, a
pairwise match stage that compares nonce pairs against a matching factor
threshold to produce Table-2, and finally a sequential disk write phase that
flushes the completed plot file to storage. The entire pipeline runs
multi-threaded on CPU using standard POSIX threads, with all stages
parallelized across available cores. As shown in Fig.~\ref{fig:cpu_results},
benchmarking was conducted on a 64-core server-grade machine, demonstrating
that plotting time decreases consistently with thread count, dropping from
44.80 minutes on a single thread to 0.87 minutes at 384 threads for a K=31
plot. Compared against the Chia reference plotter \cite{b_chia}, VaultX
achieved 4 to 50x faster plot generation depending on thread count.

% ===============================================================
%  PART 3 — DESCRIPTION OF YOUR SOLUTION
% ===============================================================
\section{Design}

This section describes the design and implementation of VaultxGPU, a
GPU-accelerated Proof-of-Space plotter built on top of the VaultX CPU
baseline. We first present the overall pipeline architecture and how
computation is distributed between the GPU and CPU, followed by a detailed
description of the CUDA implementation targeting NVIDIA hardware and the SYCL
implementation targeting AMD and Intel GPUs. Both backends share a common
Blake3 hashing kernel and produce byte-compatible plot files, differing only
in how they handle memory management, atomic operations, and key storage.

\subsection{VaultxGPU Architecture}

\begin{figure}[htbp]
    \centerline{\includegraphics[width=\figwidth]{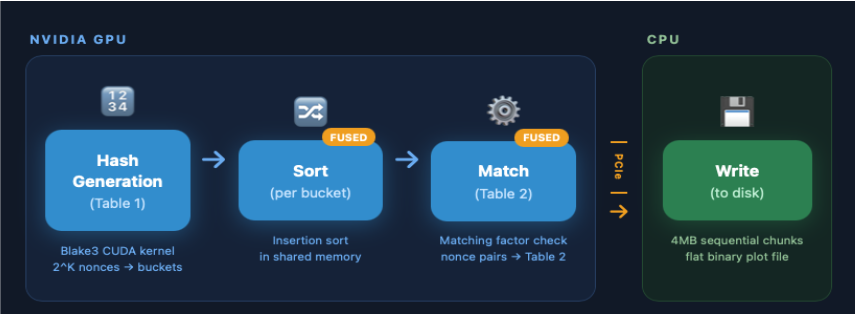}}
    \caption{VaultxGPU pipeline: Hash Generation and fused Sort+Match
    stages run entirely on the GPU in VRAM, with Table-2 transferred
    over PCIe to the CPU for disk write in 4MB sequential chunks.}
    \label{fig:gpu_pipeline}
\end{figure}

VaultxGPU redesigns the VaultX plotting pipeline to offload the three
compute-intensive stages entirely to the GPU, as illustrated in
Fig.~\ref{fig:gpu_pipeline}. In the hash generation stage, a custom Blake3
CUDA kernel launches one thread per nonce across the full $2^k$ nonce space,
hashing each nonce independently and inserting the resulting nonce-hash pair
into its corresponding bucket in GPU VRAM using atomic operations, producing
Table-1 entirely on-device without any intermediate disk write. The sort and
match stages are fused into a single GPU kernel, where each block is assigned
one bucket, loads its nonces into shared memory, performs an insertion sort by
hash value in shared memory, and then runs pairwise matching against an
expected distance threshold to produce Table-2 records, with data never
leaving shared memory between the two stages. Once Table-2 is complete, it is
transferred from GPU VRAM to the host CPU over PCIe and written to disk in
4MB sequential chunks as a flat binary plot file. This design keeps all
compute-heavy work on the GPU and limits PCIe traffic to a single transfer of
the final output, minimizing the overhead of the CPU-GPU boundary.

\subsection{CUDA Implementation (NVIDIA)}

The CUDA implementation of VaultxGPU targets NVIDIA hardware and is compiled
using \texttt{nvcc} against the NVIDIA microarchitecture directly for maximum
hardware-specific optimization. The 256-bit plot key is stored in
\texttt{\_\_constant\_\_} memory via \texttt{cudaMemcpyToSymbol}, which places
it in the GPU's broadcast cache and serves it to all threads in a warp at zero
memory bandwidth cost. Table-1 generation launches one thread per nonce with a
fixed block size of 256 threads, where each thread independently converts its
global ID to a little-endian nonce, computes a Blake3 keyed hash, extracts the
bucket index from the hash prefix, and atomically claims a slot in the
corresponding bucket using \texttt{atomicAdd}, with memory ordering, scope,
and address space all handled implicitly by the CUDA runtime. The fused
Sort+Match kernel launches one block per bucket across all $2^{24}$ buckets,
dynamically sizing shared memory per block as $\text{RPB} \times
(\texttt{NONCE\_SIZE} + \texttt{HASH\_SIZE} + 8)$ bytes to hold nonces,
hashes, and 64-bit sort keys entirely in shared memory. Following the sort and
match phase, Table-2 is streamed to disk using a 256-MB pinned host staging
buffer allocated via \texttt{cudaMallocHost}, which enables faster DMA
transfers during the PCIe write phase compared to pageable host memory.

\subsection{SYCL Implementation (AMD/Intel)}

The SYCL implementation targets AMD and Intel GPUs using Intel oneAPI DPC++
and is designed to be portable across GPU vendors while also supporting a
transparent fallback to host CPU execution when no compatible GPU device is
detected. Unlike the CUDA backend, SYCL provides no \texttt{\_\_constant\_\_}
memory abstraction; the plot key is instead allocated on the device using
\texttt{sycl::malloc\_device} and passed as a pointer argument to each kernel,
introducing an additional global memory fetch per kernel invocation compared to
the CUDA constant memory broadcast. Atomic bucket insertions require explicit
declaration of a fully-typed \texttt{sycl::atomic\_ref} object with defined
\texttt{memory\_order::relaxed}, \texttt{memory\_scope::device}, and
\texttt{access::address\_space::global\_space} parameters, in contrast to the
implicit atomics of the CUDA runtime. The SYCL backend also enforces a
per-allocation size check against \texttt{max\_mem\_alloc\_size} before
attempting to allocate Table-1 and Table-2, as some drivers, including Intel
Arc, cap individual allocations at approximately 4-GB regardless of total
available VRAM. The fused Sort+Match kernel mirrors the CUDA structure using
SYCL local memory accessors in place of CUDA shared memory, and
\texttt{sycl::group\_barrier} in place of \texttt{\_\_syncthreads()}, with
the disk write phase using a standard heap-allocated 256-MB staging buffer
rather than pinned memory, as SYCL provides no direct equivalent to
\texttt{cudaMallocHost}.

% ===============================================================
%  PART 4 — RESULTS
% ===============================================================
\section{Evaluation}

This section evaluates the performance of VaultxGPU across K-values 27
through 31, comparing the CUDA and SYCL GPU backends against both a naive
single-threaded CPU baseline and the best multi-threaded CPU configuration.
We measure total plotting time, per-stage pipeline breakdown, parallel
efficiency, and scaling behavior as problem size doubles with each K-step,
with the goal of characterizing where GPU acceleration provides the most
benefit and where bottlenecks remain.

% \subsection{Experimental Setup}
% [FILL IN]
% Testbeds: 

% CUDA:

% Mystic Eightsocket: 8 32GB VRAM Tesla V100 NVIDIA GPUs
% Mystic Epycbox: 8 16GB VRAM Tesla V100 NVIDIA GPUs
% 192-core/64-core 770GB RAM Servers.

% SYCL:
% AMD/ATI VEGA 20 Radeon GPU 16GB VRAM
% Intel DG2 Arc A770 GPU 16GB VRAM
% 6-core 48GB RAM Thermaltake PC

\subsection{Experimental Setup}

VaultxGPU was evaluated across two sets of testbeds corresponding to the
CUDA and SYCL backends respectively. For the CUDA backend, experiments were
conducted on two server-grade machines: the Mystic Eightsocket, equipped with
eight 32GB VRAM Tesla V100 NVIDIA GPUs and a 192-core CPU with 770GB RAM, and
the Mystic Epycbox, equipped with eight 16GB VRAM Tesla V100 NVIDIA GPUs and
a 64-core CPU with 770GB RAM. For the SYCL backend, experiments were conducted
on a 6-core Thermaltake PC with 48GB RAM, testing two GPU configurations: an
AMD/ATI Vega 20 Radeon GPU with 16GB VRAM and an Intel DG2 Arc A770 GPU with
16GB VRAM. The CPU baseline results were obtained on the 64-core server-grade
machine running the naive VaultX CPU plotter with thread counts ranging from 1
to 384. All implementations were evaluated across $K$-values 27 through 31,
with each $K$-value doubling the problem size relative to the previous,
and best-run times recorded across multiple runs for each configuration.

\subsection{Overall Speedup at $K=31$}

\begin{figure}[htbp]
    \centerline{\includegraphics[width=\figwidth]{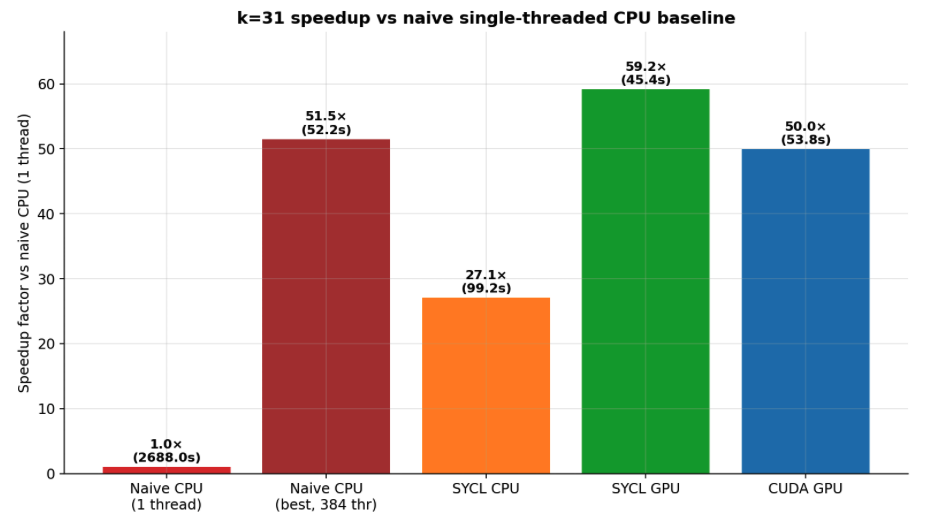}}
    \caption{Speedup at $K=31$ relative to naive single-threaded CPU
    baseline (2688s) across all implementations.}
    \label{fig:speedup}
\end{figure}

Fig.~\ref{fig:speedup} presents the speedup of all implementations at $K=31$
relative to the naive single-threaded CPU baseline of 2688 seconds. The SYCL
GPU achieves the best result at 59.2$\times$ speedup (45.4 seconds), followed
by CUDA GPU at 50.0$\times$ (53.8 seconds). The best 384-thread CPU
configuration achieves 51.5$\times$ (52.2 seconds), comparable to CUDA GPU
but at significantly higher hardware cost. The SYCL CPU fallback trails at
27.1$\times$ (99.2 seconds), nearly double the runtime of either GPU
implementation, confirming that the GPU-CPU performance gap widens as problem
size increases.

\subsection{Pipeline Stage Analysis}

\begin{figure}[htbp]
    \centerline{\includegraphics[width=\figwidth]{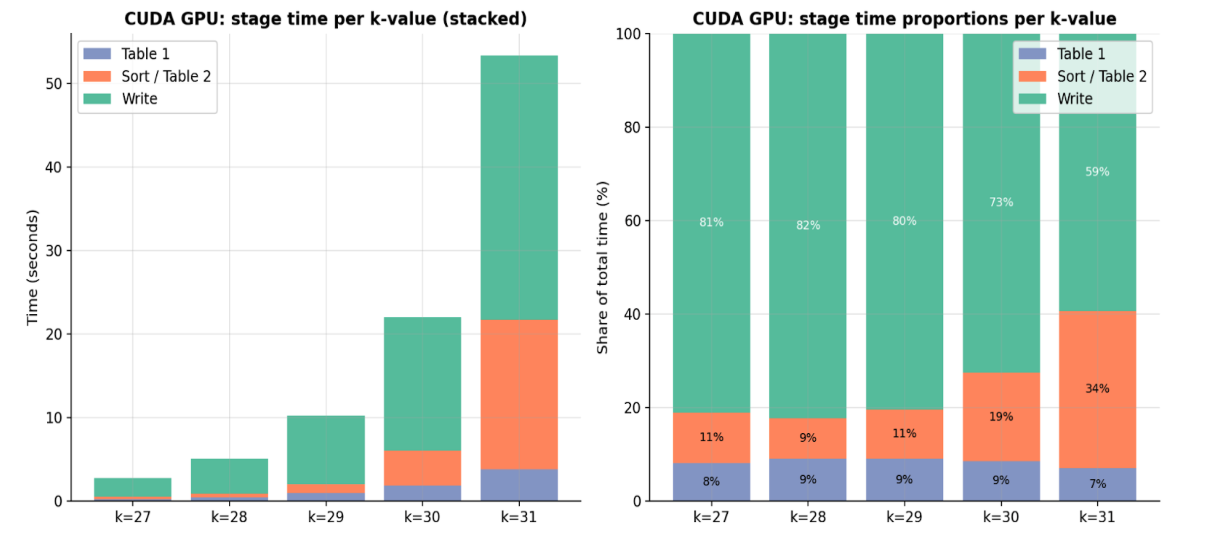}}
    \caption{CUDA GPU per-stage time breakdown across $K$-values 27--31,
    showing absolute runtimes (left) and proportional stage shares (right).}
    \label{fig:io_bottleneck}
\end{figure}

Fig.~\ref{fig:io_bottleneck} presents the per-stage time breakdown of the
CUDA GPU pipeline across $K$-values 27 through 31. The disk write stage
dominates across all $K$-values, accounting for 80--82\% of total time at
$K=27$--$29$ and remaining the largest stage at $K=31$ with 59\%. The
Sort/Table-2 stage grows from 11\% at $K=27$ to 34\% at $K=31$, reflecting
the $O(n^2)$ insertion sort bottleneck as bucket sizes increase. Table-1
generation remains the smallest stage at 7--9\% across all $K$-values,
confirming that the Blake3 hashing kernel scales efficiently with problem
size.

\subsection{CPU vs.\ GPU Parallel Efficiency}

\begin{figure}[htbp]
    \centerline{\includegraphics[width=\figwidth]{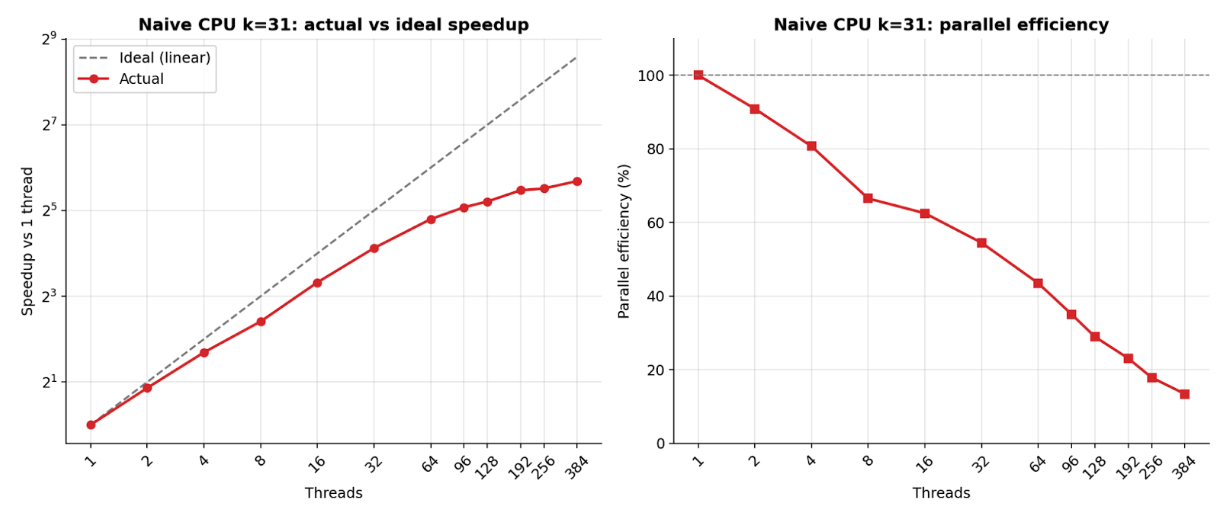}}
    \caption{Naive CPU $K=31$ actual vs.\ ideal speedup (left) and parallel
    efficiency (right) across thread counts 1--384.}
    \label{fig:cpu_vs_gpu}
\end{figure}

Fig.~\ref{fig:cpu_vs_gpu} shows the actual versus ideal speedup and parallel
efficiency of the naive CPU implementation at $K=31$. While speedup improves
consistently from 1 to 384 threads, it diverges sharply from ideal linear
scaling beyond 16 threads, with parallel efficiency dropping from 100\% at 1
thread to just 14\% at 384 threads. The best CPU result of 52.2 seconds at
384 threads remains slower than the SYCL GPU at 45.4 seconds, confirming that
additional CPU threads cannot close the performance gap with GPU acceleration.

\subsection{Scaling with Problem Size}

\begin{figure}[htbp]
    \centerline{\includegraphics[width=\figwidth]{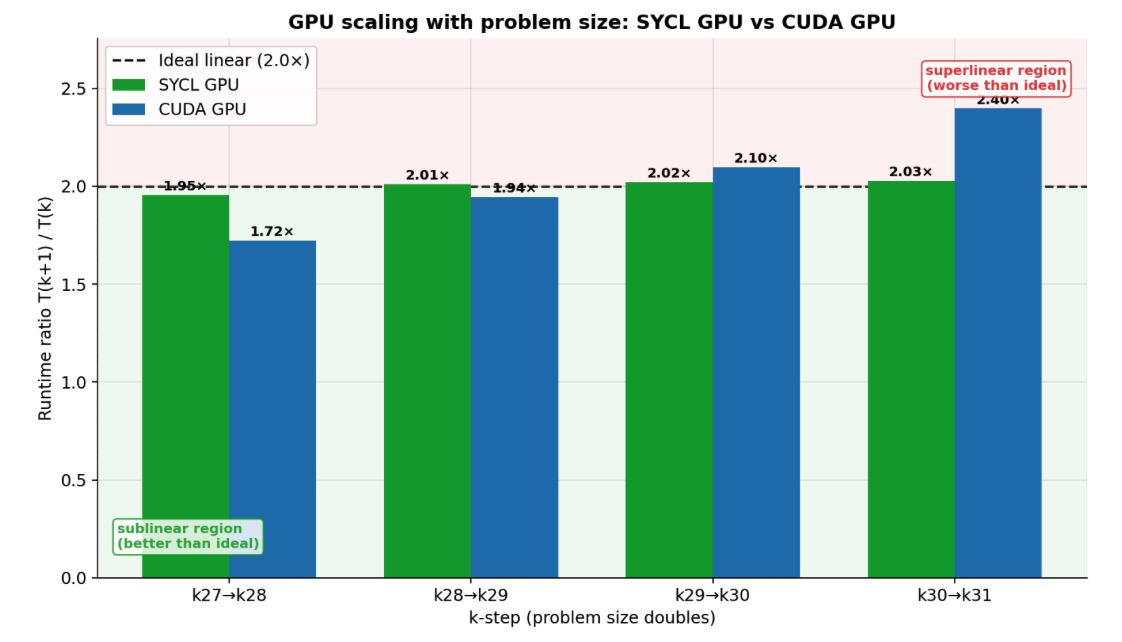}}
    \caption{GPU scaling with problem size: runtime ratio $T(k+1)/T(k)$
    for SYCL GPU and CUDA GPU across $K$-steps 27--31, with ideal
    linear scaling at 2.0$\times$.}
    \label{fig:scaling}
\end{figure}

Fig.~\ref{fig:scaling} presents the runtime ratio $T(k+1)/T(k)$ for both GPU
implementations as problem size doubles with each $K$-step, with an ideal
linear scaling ratio of 2.0$\times$ shown as a reference. The SYCL GPU
maintains near-ideal linear scaling across all $K$-steps, with ratios of
1.95$\times$, 2.01$\times$, 2.02$\times$, and 2.03$\times$ from $K=27$
through $K=31$, demonstrating that SYCL scales predictably as problem size
grows. The CUDA GPU begins in the sublinear region at $K=27$$\rightarrow$$K=28$
with a ratio of 1.72$\times$, converges toward ideal at
$K=28$$\rightarrow$$K=29$ with 1.94$\times$, and then enters the superlinear
region at $K=30$$\rightarrow$$K=31$ with a ratio of 2.40$\times$, indicating
that CUDA kernel management becomes increasingly inefficient at larger bucket
sizes and higher $K$-values.

\subsection{Cross-Implementation Runtime Comparison}

\begin{figure}[H]
    \centerline{\includegraphics[width=\figwidth]{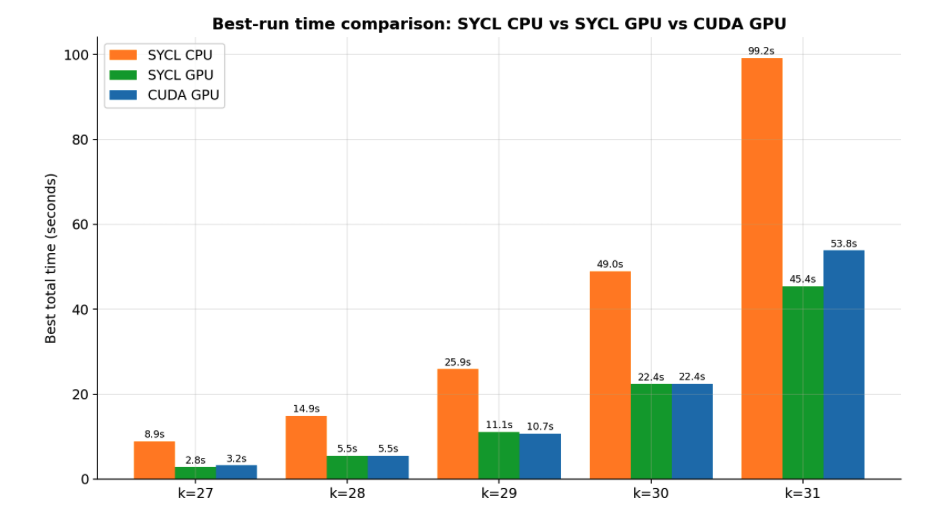}}
    \caption{Best-run time comparison across all $K$-values for SYCL CPU,
    SYCL GPU, and CUDA GPU implementations.}
    \label{fig:best_results}
\end{figure}

Fig.~\ref{fig:best_results} compares the best total runtime across all
$K$-values for SYCL CPU, SYCL GPU, and CUDA GPU implementations. Both GPU
implementations consistently outperform the SYCL CPU across all $K$-values,
with the performance gap widening significantly as $K$ increases. At $K=27$
the SYCL CPU completes in 8.9 seconds compared to 2.8 seconds for SYCL GPU
and 3.2 seconds for CUDA GPU. By $K=31$, the SYCL CPU runtime nearly doubles
to 99.2 seconds while SYCL GPU and CUDA GPU reach only 45.4 seconds and 53.8
seconds respectively, demonstrating that GPU runtimes scale far more
gracefully than CPU runtimes as problem size grows. At $K=30$, SYCL GPU and
CUDA GPU achieve identical runtimes of 22.4 seconds, with the gap between the
two opening only at $K=31$ as CUDA kernel management degrades at larger bucket
sizes.

% ===============================================================
%  PART 5 — ANALYSIS AND CONCLUSION
% ===============================================================
\section{Analysis and Conclusion}

VaultxGPU demonstrates that GPU acceleration is a viable and effective
direction for Proof-of-Space plot generation. The SYCL GPU implementation
achieves a 59.2$\times$ speedup over a single-threaded CPU baseline at
$K=31$, completing a plot in 45.4 seconds compared to 2688 seconds, and
outperforms even the best 384-thread CPU configuration at a significantly
lower hardware cost. The CUDA GPU implementation achieves 50.0$\times$
speedup, confirming that both NVIDIA and AMD hardware can effectively
accelerate the Blake3 hashing pipeline. These results establish GPU
acceleration as the correct direction for scaling Proof-of-Space plotting
beyond what CPU parallelism can achieve.

The most effective design decision was fusing the sort and Table-2 match
stages into a single GPU kernel operating entirely within shared memory,
eliminating any intermediate global memory round-trip between the two stages.
SYCL's portability across AMD and Intel hardware also proved to be a
significant advantage, allowing a single codebase to target multiple GPU
vendors with minimal changes. However, two design decisions did not perform
as well as anticipated. The insertion sort running on a single thread per
bucket introduced an $O(n^2)$ bottleneck that became increasingly significant
at larger $K$-values, contributing to Sort/Table-2 growing from 11\% of
total pipeline time at $K=27$ to 34\% at $K=31$. Additionally, CUDA kernel
management degraded at larger bucket sizes, causing CUDA to fall into the
superlinear scaling region at $K=30$$\rightarrow$$K=31$ with a runtime ratio
of 2.40$\times$, well above the ideal 2.0$\times$.

The most significant bottleneck identified across all $K$-values was disk I/O
during the write phase, accounting for 59--82\% of total pipeline time
depending on $K$-value. This bottleneck persists regardless of GPU performance
because the write stage runs entirely on the host CPU and is inherently
sequential. If we were to redo this project, we would address this from the
start by overlapping PCIe transfers with GPU computation using asynchronous
CUDA streams, and by investigating direct NVMe write paths to reduce host CPU
involvement in the write pipeline. We would also replace the single-threaded
insertion sort with a GPU optimized parallel sort algorithm such as radix sort
or merge sort to eliminate the $O(n^2)$ bottleneck, and profile CUDA kernel
occupancy earlier in development to close the performance gap with the SYCL
implementation.

Future work will focus on three directions. First, replacing the insertion
sort with a GPU-parallel sort algorithm to eliminate the $O(n^2)$ bottleneck
that becomes the dominant cost at larger $K$-values. Second, optimizing the
disk write pipeline by overlapping PCIe transfers with GPU computation and
exploring direct NVMe write paths to reduce the 59--82\% I/O overhead that
currently dominates the pipeline. Third, extending VaultxGPU to support
$K$-values beyond 31 through chunked VRAM tiling, which would allow the
plotter to handle problem sizes that exceed the physical VRAM capacity of a
single GPU without falling back to CPU execution.

% -------------------------------------------------------
%  ACKNOWLEDGMENT
% -------------------------------------------------------
\end{doublespace}

\section*{Acknowledgment}
The authors thank Dr.\ Luke Logan for supervision and the Illinois Institute
of Technology College of Computing for providing the research infrastructure
used in this project.

\section*{Author Contributions}
Om Amit Gandhi designed and implemented the CUDA backend targeting NVIDIA hardware,
including the Blake3 hashing kernel, fused Sort+Match kernel, and pinned-memory
disk write pipeline. Samuel Fatunmbi implemented the SYCL backend targeting AMD
and Intel GPUs, conducted all benchmark experiments across K-values 27 through 31,
and produced the performance figures and analysis presented in Section IV.
Both authors contributed equally to the overall system architecture and the writing
of this paper.

% -------------------------------------------------------
%  REFERENCES
% -------------------------------------------------------
\bibliographystyle{IEEEtran}
\bibliography{references}

\end{document}